\documentclass[acmsmall]{acmart}

\usepackage{array,multirow,etoolbox,graphicx,subcaption,fancyhdr,tabularx,longtable}
\usepackage{siunitx}
\usepackage[version=4]{mhchem}
\usepackage{booktabs,makecell}

\newcounter{rowcntr}[table]
\renewcommand{\therowcntr}{\arabic{rowcntr}}
\newcolumntype{N}{>{\refstepcounter{rowcntr}\therowcntr}c}
\AtBeginEnvironment{tabular}{\setcounter{rowcntr}{0}}

\setcopyright{rightsretained}


\setcopyright{rightsretained}
\acmJournal{TOSEM}

\newcommand{\add}[1]{\textcolor{blue}{#1}}

\begin{document}
\author{Chandra Maddila}
\additionalaffiliation{%
\institution{Microsoft Research}
}
\affiliation
{
\institution{Microsoft Research}
\city{Redmond}
\state{WA, USA}
}
\email{chandu.maddila@gmail.com}

\author{Sai Surya Upadrasta}
\affiliation
{
\institution{Microsoft Research}
\city{Bengaluru}
\state{India}
}
\email{upadrastasaisurya1@gmail.com}

\author{Chetan Bansal}
\affiliation
{
\institution{Microsoft Research}
\city{Redmond}
\state{WA, USA}
}
\email{chetanb@microsoft.com}

\author{Nachiappan Nagappan}
\additionalaffiliation{%
\institution{Microsoft Research}
}
\affiliation
{
\institution{Microsoft Research}
\city{Redmond}
\state{WA, USA}
}
\email{nachiappan.nagappan@gmail.com}

\author{Georgios Gousios}
\affiliation
{
\institution{Delft University of Technology}
\city{Delft}
\state{The Netherlands}
}
\email{g.gousios@tudelft.nl}

\author{Arie \lowercase{van} Deursen}
\affiliation
{
\institution{Delft University of Technology}
\city{Delft}
\state{The Netherlands}
}
\email{arie.vandeursen@tudelft.nl}

\renewcommand{\shortauthors}{C. Maddila, S. Upadrasta, C. Bansal, G. Gousios, and A. Van Deursen}
\newcommand{\plr}{Pull request}
\newcommand{\csharp}{C\#}
\newcommand{\prl}{PullRequest Lifetime}

\newcommand{\todo}[1]{\textbf{\textcolor{red}{TODO: #1 }}}

\newcommand{\pr}{pull request}
\newcommand{\prc}{Pull request}
\newcommand{\PaperTitle}{Nudge}
\newcommand{\ado}{Azure DevOps}
\newcommand{\NumRepos}{147}
\newcommand{\NumNudgedPrs}{8,500}

\newcommand{\commentblock}[1]{}
\begin{abstract}
\prc{}s are a key part of the collaborative software development and code review process today. However, \pr{}s can also slow down the software development process when the reviewer(s) or the author do not actively engage with the \pr{}. In this work, we design an end-to-end service, Nudge, for accelerating overdue \pr{}s towards completion by reminding the author or the reviewer(s) to engage with their overdue \pr{}s. First, we use models based on effort estimation and machine learning to predict the completion time for a given \pr{}. Second, we use activity detection to filter out pull requests that may be overdue, but for which sufficient action is taking place nonetheless. Lastly, we use actor identification to understand who the blocker of the \pr{} is and \textit{nudge} the appropriate actor (author or reviewer(s)). The~key novelty of Nudge is that it succeeds in reducing pull request resolution time,
while ensuring that developers perceive the notifications sent as useful,
at the scale of thousands of repositories.
In a randomized trial on \NumRepos{} repositories in use at Microsoft,
Nudge was able to reduce pull request resolution time by 60\% for \NumNudgedPrs{} pull requests,
when compared to overdue pull requests for which Nudge did not send a notification.
Furthermore, developers receiving Nudge notifications resolved 73\% of these notifications as positive.
We~observed similar results when scaling up the deployment of Nudge to 8,000 repositories at Microsoft,
for which Nudge sent 210,000 notifications during a full year.
This demonstrates Nudge's ability to scale to thousands of repositories.
Lastly, our qualitative analysis of a selection of Nudge notifications indicates areas for future research, such as taking dependencies among pull requests and developer availability into account.
\end{abstract}

\title{Nudge: Accelerating Overdue Pull Requests Towards Completion}
\maketitle
\renewcommand{\shortauthors}{C. Maddila et al.}

\section{Introduction}
With the adoption of collaborative software development platforms like GitHub and Azure DevOps, \pr{}s have become the standard mechanism for distributed code reviews. \prc{}s enable developers as automated agents to collaboratively review the code before it gets integrated into the mainline development.
Once the reviewers have signed off on the changes these can be merged with the main branch and deployed. \prc{}s has recently become an active area of research in the software engineering community.
Various aspects of \pr{}s have been studied, such as reviewer recommendation \cite{asthana2019whodo, yu2016reviewer}, prioritization \cite{van2015automatically}, and duplication \cite{wang2019duplicate}.
Additionally, several bots and extensions have been built for platforms like GitHub and Azure DevOps to automate various software development workflows \cite{dey2020detecting, kumar2019building}.

While \pr{}s streamline the code review process significantly, they can also slow down the software development process. For instance, if the reviewers are overloaded and lose track of the \pr{}, it might not be reviewed in a timely manner. Similarly, if the \pr{} author is not actively working on the \pr{} and reacting to the reviewers' comments, the review process could be slowed down significantly.
Hence, if the \pr{}'s author and reviewers do not actively engage, the \pr{}s can remain open for a long time, slowing down the coding process and possibly causing side effects such as merge conflicts.
Yu et al. \cite{Yu:2015:WDP:2820518.2820564} did a retrospective study of the factors impacting \pr{} completion times.
They found that pull request latency requires many independent variables to explain adequately,
with the size of the \pr{} and the presence of a continuous integration pipeline as major factors.
Long-lived feature branches can also cause several unintended consequences \cite{NewRelic}. Some of the most common side effects caused by long-lived feature branches or \pr{}s are:
\begin{itemize}
    \item They hinder communication. \prc{}s that are open for longer periods of time hide a developer's work from the rest of their team. Making code changes and merging them quickly increases source code re-usability by making the functionality and optimizations built by a developer available to other developers. 
    \item 
    In large organizations with thousands of developers working on the same codebase, the assumptions that a developer may make about the state of the code might not hold true, the longer they have their feature branches open. The developers become unaware of how their work affects others.
    \item 
    Long-lived pull requests cause integration pain. When the code is merged more frequently to the main branch, integration testing can be done earlier, issues can be detected faster and bugs can be fixed at the earliest possible moment.
    \item 
    Branches that stay diverged from the main branch for longer periods of time can cause complex merge conflicts that are hard to solve. Dias et al. \cite{dias2020understanding} studied over 70,000 merge conflicts and found that code changes with long check-in times are more likely to result in merge conflicts.
    \item 
    Overdue pull requests prevent companies from delivering value to their customers quickly.
    Organizations can deliver more value to their stakeholders by releasing new features or bug fixes 
    in the organization's products or services earlier if the corresponding code is merged faster.
\end{itemize}

In order to address these concerns, we designed and deployed Nudge, a service for accelerating overdue \pr{}s towards completion. As its name suggests, Nudge sends a reminder if a pull request is overdue. We carefully designed Nudge so that it (1) actually achieves faster pull request resolution; (2) minimizes the number of notifications it sends to avoid disturbing developers unnecessarily; and (3) can operate at the scale of thousands of repositories and developers.

To realize these objectives, Nudge relies on effort estimation to predict the completion time for a given \pr{}.
Next, it determines activities and identifies the actor (the reviewer(s) or the author) blocking the \pr{} from completion.
It then notifies the identified actor through the comment functionality of the \pr{} environment.

To design and build Nudge, we first perform correlation analysis to understand which factors impact \pr{} completion time.
We look at factors related to the \pr{}, its author, the underlying system, the team, and the role of the developer in the team.
Unlike Yu et al. \cite{Yu:2015:WDP:2820518.2820564}, we only consider factors that are known at the time of the \pr{} creation.

Next, we use effort estimation for predicting the \pr{} completion time at the time of \pr{} creation. Effort estimation models have been long studied in software engineering research. We build a model for predicting the completion time of a \pr{} on the rich body of work in the effort estimation literature.
Prior work \cite{inproceedings} has focused on effort estimation at the feature and project level, but not at the level of individual \pr{}s. We use several metrics from the defect prediction literature like code churn \cite{Ostrand:2004:BUG:1007512.1007524}, reviewer information \cite{7950877}, and ownership information \cite{inproceedings} to build our \pr{} lifetime prediction model.

While effort estimation models have been shown to be accurate \cite{attarzadeh2012proposing}, they cannot account for contextual and environmental factors such as workload of the \pr{} reviewer(s) of the author. Therefore to improve the notification precision, we implement \emph{activity detection} which monitors any updates on the \pr{}, such as new commits or review comments, and adjusts the notification accordingly.
Furthermore, to determine \emph{who} needs to receive the notification,
we implement \emph{actor identification} to infer the actor (\pr{} author or specific reviewer(s)) who is blocking the \pr{} from completion.

To assess to what extent Nudge has been able to meet its objectives, 
we conducted a number of experiments.
To assess pull request resolution time and developer perception of notifications,
we deployed Nudge to \NumRepos{} repositories, using its telemetry functionality to collect data for a period of nine months.
During this period, Nudge identified 12,356 pull requests that were taking longer than the time Nudge predicted.
We employed Nudge via a randomized trial by sending a notification to a subset of \NumNudgedPrs{} (55\%) randomly selected pull requests,
thus allowing us to compare their resolution time with those for which no notification was sent.
Our findings indicate a reduction of 60.62\% in average \pr{} lifetime thanks to the use of Nudge.
The vast majority (81.53\%) of the notified pull requests are closed within a week.

To be able to assess the developer's perception of the Nudge notifications, we give users of Nudge recommendations the option to provide
feedback, both via a negative/neutral/positive tick box and an open text field.
We find that 73\% of the pull requests received a positive resolution from the developers.
We used the open answers to identify areas for future improvements, such as taking dependencies between pull  requests into account (in case one pull request is blocking another).

To assess the scalability of Nudge, we monitored its deployment on 8,000 different systems at Microsoft from January 2021 until December 2021.
During this period, Nudge sent 210,000 notifications authored by 40,000 unique developers.
Since this is an actual deployment, unlike the randomized trial in our experiments, we have no ``untreated'' data points to compare to.
Nevertheless, we see that 83.65\% of the nudged pull requests are closed within a week, which is consistent with the findings from the randomized trial.
Also, user satisfaction is similar, with 71\% of the notifications receiving a positive resolution.
From this, we conclude that the design of Nudge permits operation at the scale of thousands of repositories and that the positive results in terms of time reduction and user satisfaction remain valid.

Thus, the novelty of this paper lies in the following key contributions:
\begin{enumerate}
    \item 
    We propose a novel approach to warning developers and reviewers of pull requests when they are running late,
    combining effort estimation, activity detection, and actor identification (Sections~\ref{SystemDesign}--\ref{sec:status}).
    \item
    We design and deploy a scalable implementation of our approach in a tool called Nudge (Section~\ref{Implementation}).
    \item
    We demonstrate that the use of Nudge leads to a 60\% speed-up of delayed pull requests and that over 70\% of the developers warned about their pull requests appreciate such warnings as positive (Section~\ref{sec:eval}).
    \item
    We apply Nudge to 8,000 systems and demonstrate that its benefits remain present at scale (Section~\ref{sec:eval}).
\end{enumerate}

This paper is a substantially revised extension of our earlier publication~\cite{madillaFSE}.
New in the present paper is the use of activity detection and actor identification, the evaluation of these, and the discussion of
the application  of Nudge to 8,000 systems in the period January--December 2021.

\section{Related Work} \label{RelatedWork}

Our research relies on effort estimation techniques to determine the amount of time 
needed to decide whether a given pull request can be merged.
Software effort estimation is a field of software engineering research that has been studied 
extensively in the past four decades \cite{Boehm:1984:SEE:2090.2283139, Briand:2000:ERD:347260.347278, Chulani:1999:BAE:322677.322688, 6363444, Bettenburg:2015:TIS:2769781.2769787}. 
Typically, in this line of research, one tries to predict either the effort needed to complete the entire project or the effort needed to finish a feature.
One of the earliest effort estimation models was the COCOMO model proposed by Barry W. Boehm in his 1981 book, Software Engineering Economics \cite{Boehm:1984:SEE:2090.2283139}, which he later updated to COCOMO 2.0 in 1995 \cite{article}. This work was followed up by Briand et al. \cite{Briand:1999:ACC:302405.302647} who compared various effort estimation modeling techniques using the data set curated by the European Space Agency. In all these cases a model was built for the entire software project and effort was estimated for function points. More recently, Menzies et al. \cite{6363444} and Bettenburg et al. \cite{Bettenburg:2015:TIS:2769781.2769787} looked at the variability present in the data and therefore built separate models for subsets of the data.

More recently there has been interest in predicting pull request acceptance,
both the eventual decision (merge or abort), as well as the time, needed to make the decision.
Soares et al. \cite{Soares:2015:AFP:2695664.2695856}, and Tsay et al. \cite{Tsay:2014:IST:2568225.2568315} looked at a variety of factors to see which one had an impact on pull request acceptance. More specifically, Terrell et al. \cite{10.7717/peerj-cs.111} and Rastogi et al. \cite{Rastogi:2018:RGL:3239235.3240504} looked at gender or geographical location impact on a pull request acceptance. The work closest to our work is by Yu et al. \cite{Yu:2015:WDP:2820518.2820564}, who explored the various factors that could impact how long it took for an integrator to merge a pull request. Unlike their study, we do not examine what factors might impact the time taken to accept a pull request, but rather how much time it would actually take for a pull request to be accepted. Hence, unlike past papers which were empirical studies on building knowledge with respect to pull request acceptance, we build a system that will predict how long it will take to accept a pull request and provide actionable feedback to the developers leveraging that knowledge. 

As shown by various studies \cite{Boehm:1984:SEE:2090.2283139, Briand:2000:ERD:347260.347278, Chulani:1999:BAE:322677.322688, 6363444, Bettenburg:2015:TIS:2769781.2769787}, effort estimation is a hard problem. One of the primary reasons that contribute to the errors is changing organizational dynamics, the landscape of the competition, and ever-changing schedules, and priorities. Doing effort estimation at the \pr{} level reduces the uncertainty and the variability up to some degree.

Our ambition to devise a technique to warn developers about pull request delays can be viewed as a software development bot.
The extensibility mechanisms provided by software development platforms like GitHub and Azure DevOps have enabled a huge ecosystem \cite{GitHubMarketplace} of bots and automated services. It has also spawned active research \cite{storey2020botse} on understanding and building bots to assist with various software engineering tasks. Storey et al. \cite{storey2020botse} have defined a \emph{Software Engineering Bot} as software that automates a feature, performs a function normally done by humans, and interacts with humans. Lebeuf et al. \cite{lebeuf2019defining} have proposed a taxonomy for software bots based on the environment, the intrinsic properties of the bot, and the bot's interaction with the environment. In terms of applications, prior work has focused on improving the code review process by automating reviewer recommendation \cite{yu2016reviewer, asthana2019whodo}, diagnosing issues \cite{mehta2020rex, bhagwan2018orca, bansal2020decaf}, refactoring \cite{ren2019identifying, wyrich2019towards}, and even intent understanding of the code changes \cite{wang2020large, wang2019leveraging}. In this work, we  built and deployed Nudge which is a bot for increasing software development velocity and productivity by accelerating PR completion.

Bots warning about potential delays can be found beyond software development systems.
General workflow management systems such as IFTTT (if-this-then-that)  \cite{IFTTT}
and Microsoft Power Automate \cite{PowerAutomate} can be used for creating various automation workflows in domains such as smart home automation \cite{KUMER20214070}, healthcare \cite{9391828}, and smart mobility \cite{MENON2020100213}.
One of the biggest challenges with such tools is that they cannot take into account the complexity associated with internal state changes, and interactions between various actors in the systems they operate on.
Such general systems are well suited for tasks like sending daily reports or reminders based on simple logic. For example, a \pr{} reminder system that is built using Power Automate \cite{PowerAutomatePR} could check if a \pr{} is active. If it is active, the system can trigger an email, on a pre-defined cadence, to all the reviewers of the \pr{}.
While technically speaking such general notification systems could play a role in the implementation of Nudge, we offer tight integration with the pull request environment instead.
This is not only most natural to the developers involved, but also enables us to
determine the various attributes and state changes happening in each \pr{} based on which an alert has to be triggered, as well as the branching conditions which help determine whom the notification should be redirected to.

\section{Background: A Pull Request's Life Cycle}
\label{Background}



\begin{figure}
\centering
\includegraphics[width=0.75\columnwidth]{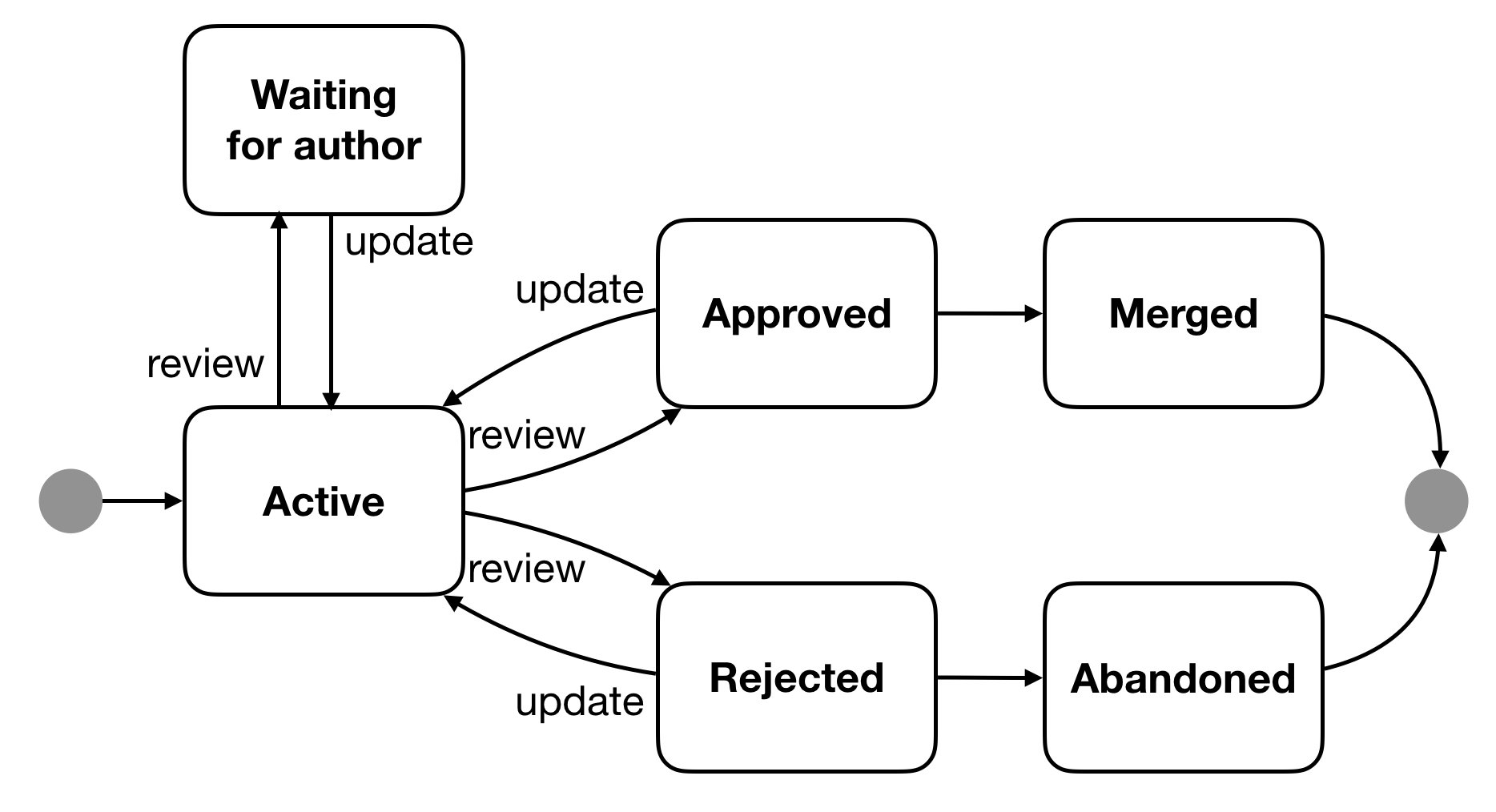}
\vspace{-0.5\baselineskip}
\caption{The life cycle of a \pr{}}
\label{PrStateDiagram}
\end{figure}

In this paper, we assume a pull request goes through the life cycle as depicted in Figure~\ref{PrStateDiagram}. Based on this, a pull request can be in one of the following states:

\begin{description}
    \item[Active]
        The pull request has been published by the author.
        Reviewers are assigned and the \pr{} is open for code \emph{review}.
    \item[Waiting for author]
        The reviewer has left review comments and expects the author to \emph{update} the code to address them.
    \item[Approved]
        The reviewer was satisfied with the code changes in the \pr{} and approved it to be merged with the main branch. Thus, the author can merge and finalize, or, optionally, decide additional \emph{updates} are called for and re-start the reviewing process from the \textbf{Active} state.
    \item[Rejected]
        The reviewers are not satisfied with the code changes and reject the \pr{}.
        The author can attempt additional \emph{updates} to restart the reviewing process, but otherwise, the pull request will be rejected.
    \item[Merged]
        After the reviewers signed off on the \pr{}, the author successfully merged the code into the main branch.
    \item[Abandoned]
        The author of the \pr{} decides to not pursue the code changes further.
\end{description}
After the pull requested has been merged or abandoned, the pull  request is closed and cannot be re-opened again (developers would need to open a \emph{new} pull request instead).

To transition between these states, there are three different \emph{actors} involved:
Authors, reviewers, and non-human actors (bots):



\begin{description}
    \item[Authors:] Authors create a \pr{} in the first place. They send the \pr{} for review and keep working on the \pr{} by reacting to the review comments by pushing new changes (in the form of commits or iterations). Once all the reviewers are satisfied, they make the final decision to merge or not merge a change. They have a significant influence on the pace of the \pr{}. If they react to the review comments quickly and resolve them, \pr{} will have a better chance of making progress quickly.
    In Figure~\ref{PrStateDiagram}, they can trigger the transitions labeled with the \emph{update} event.
    
    \item[\textbf{Reviewers}] Reviewers are added by the authors or any other automation tools (based on certain conditions) to \pr{}s. Reviewers have a responsibility to perform a thorough code review and provide their feedback. The agenda of the reviewers is to ensure the quality of the source code stays high and adheres to the standards imposed by their respective teams or organizations. Reviewers can be individuals in the same team or people with more experience and expertise in the area of source code that is being changed or groups that are a collection of individual reviewers. When a \pr{} is submitted for a review, reviewers can either approve or reject or make suggestions that need to be acted upon by the author of the change and resolve the comments made by the reviewers. By virtue of their role, reviewers can significantly impact the outcome of the review and the velocity of the \pr{}.
    In Figure~\ref{PrStateDiagram}, they can trigger the \emph{review} transitions.
    
    \item[\textbf{Non-human actors}] With the increased use of bots and automation tools, non-human actors can also play a role in determining the velocity with which change progression happens. Tools that enforce security and compliance policies or styling guidelines, or that ensure dependencies are not broken are some of the examples. Such bots can place comments like a reviewer would, and thus trigger \emph{review} transitions in Figure~\ref{PrStateDiagram}. The non-human actors, which sometimes act as code reviewers, do not contribute to the time taken to review \pr{}s. However, they impact the \pr{} status determination algorithm (explained in Section \ref{ActivityDetection}) by influencing the \pr{} state changes.
\end{description}


Pull request life-cycle is a complex process involving several actors and activities. However, it's also an important process since inadequate code reviews can result in bugs and sub-optimal design with both short-term and long-term implications. Prior work has shown that the size of the code changes has a significant impact on the time taken for code reviews. However, there are several other factors that can also impact code review time. Baysal et al. \cite{baysal2013influence} found that the reviewer's workload and past experience can impact the time taken for code reviews. Further other organizational (such as release deadlines) and geographical factors (collaboration across multiple time zones) can also influence the speed. While these factors are critical for faster code reviews but they are hard to change. 

Often, developers are working on multiple projects and features at the same time. They are simultaneously working on code changes while reviewing other people's code reviews. So, it's very common to lose track of pending activities that might be blocking the pull requests. This problem is further amplified since these code reviews are spread across multiple repositories. So, in this work, we build the Nudge tool to provide intelligent reminders to both the authors and the reviewers.

\section{Nudge System Design} \label{SystemDesign}
The side effects manifested by \pr{}s that are open for longer periods of time, are prevalent in large organizations like Microsoft, as well as in large open-source projects. Because of that, there has been a demand inside such organizations for a service that can help engineering teams alleviate the problems induced by long-running \pr{}s. We designed the Nudge system to address this problem and operationalized it across \NumRepos\ repositories. We then performed a large-scale testing/validation of the effectiveness of the Nudge system by analyzing various metrics and collecting user feedback. In this section, we describe the design of the Nudge system in detail. 

\subsection{Design Overview} \label{DesignOverview}
The Nudge system consists of three main components: A machine learning-based effort estimation model that predicts the lifetime of a given \pr{}, an activity detection module to establish what the current state of the \pr{} is, and an actor determination module to identify who would be need to take action.

\paragraph{Prediction model} The Nudge system leverages a prediction model to determine the lifetime for every \pr{}. The model is a linear regression model as explained in Section~\ref{sec:prediction}. We performed the regression analysis to understand the weights of each of the features and how they impact the ability of the model to accurately predict the lifetime for a given \pr{}. 
We use historical \pr{} data to extract some of the features and the dependant variable (\pr{} lifetime). 

For the repositories where we have enough training data, i.e., at least thousands of data points (or \pr{}s), we train a repository-specific model. If the repository is small or new and it does not have many \pr{}s that is completed, we use a global model that is trained on all the repositories' data.
Once the repository matures and records enough activity, we train a repository-specific model and deploy it.
The models are retrained, through an offline process, periodically, to adjust to the changes in the feature weights and changing repository dynamics. Every time the model is retrained, we use a moving window to fetch the data from the last two years (from the date of retraining) to make sure the training data reflects the ever-changing dynamics and takes into account the changes happening to the development processes.

\paragraph{Activity Detection}
The role of the activity detection module is to help the Nudge system understand if there has been any activity performed by the author or the reviewer of the \pr{} of late. This helps the Nudge system not send a notification, even though the lifetime of the \pr{} has exceeded its predicted lifetime. This module serves as a gatekeeper that gives the Nudge system a `go' or `no go' by observing various signals in the \pr{} environment.

\paragraph{Actor Identification} The primary goal of this module is to determine the blocker of the change (the author or a reviewer) and engage them in the notification, by explicitly mentioning them. This module comes into action once the \pr{} meets the criteria set by the prediction module and the Activity Detection modules. Once the Nudge system is ready to send the notification, the Actor Identification module provides information to the Nudge notification system to direct the notification towards the change blocker.

\paragraph{Nudge Workflow} \label{NudgeWorkflow}
The three modules are combined with a notification system to form Nudge as shown in Figure~\ref{fig:WorkFlow}. This results in the following workflow:


\begin{figure}
\includegraphics[width=0.9\columnwidth]{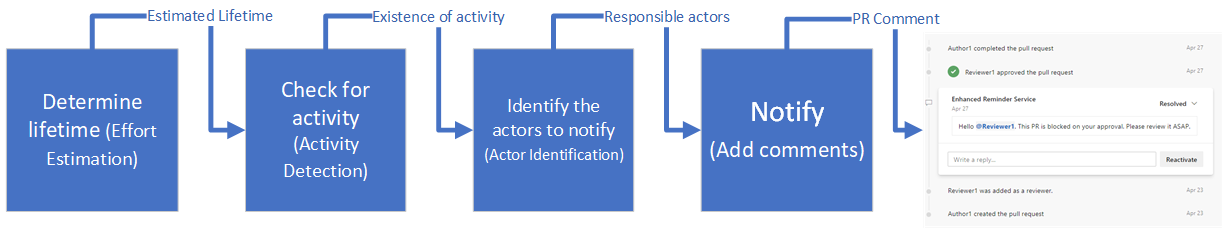}
\caption{\PaperTitle\ Workflow}
\label{fig:WorkFlow}
\end{figure}

\begin{enumerate}

    \item 
    The \PaperTitle\ service workflow starts with calculating the effort needed for a \pr{} using  effort estimation models. When the corresponding batch job is triggered, it first scans all active \pr{}s and runs the effort estimation model (see Section~\ref{sec:prediction}) to determine the lifetime of a \pr{} and save it to a back-end SQL database. The batch job is triggered every six hours. 
    \item Once a \pr{}'s actual lifetime crosses the estimated lifetime (using the effort estimation models), the next module, Activity Detection, is run which checks for any activity in the \pr{} environment. If there is an activity observed in the last 24 hours, the workflow is terminated.
    \item Once the activity detection algorithm determines that there was no activity in the last 24 hours, the Actor Identification algorithm kicks in which determines the change blockers and dependant actors who should take appropriate actions to facilitate the movement of the \pr{}s.
    \item Finally, notifications are sent to the list of actors identified in the previous step in the form of \pr{} review comments and email messages. By design, Nudge sends at most one notification per \pr{}.
\end{enumerate}

\subsection{Key Design Considerations}

\paragraph{Feature extraction} Nudge's machine learning model needs to extract various features for every new \pr{} to perform inference. There are three classes of features that constitute the feature vectors:
\begin{enumerate}
    \item Some features are easy to extract and are readily available in the \pr{}. 
    Examples include the day of the week, the length of its description, etc.
    \item Some features can be computed based on the information available in the \pr{}. An example is whether the \pr{} is a new feature or a bug fix. 
    For this, we run the relevant heuristic algorithm to classify the \pr{}s accordingly.
    \item 
    Some features are hard to calculate on the fly as they require mining historical data.
    Examples include the average lifetime of the \pr{}s created by an author or the average lifetime of the \pr{}s that edit specific area paths in the source code.
    For these, we compute their value through a batch job that runs at a scheduled frequency (every six hours) and that stores the results in a database. 
    Upon inference, the pre-computed features are queried from the database and appended to the other two types of features to form the feature vectors.
\end{enumerate}

\paragraph{Scale}
Nudge has two primary scale challenges it has to deal with. The first is conducting feature extraction, training, and re-training for over 100 repositories. The second challenge is inference and sending notifications on live \pr{}s created in these repositories.
To deal with the first problem, we adopted a strategy to train the model by pre-computing some of the features beforehand, when the data is ingested itself. This helped in reducing the overall training time. The second strategy we adopted is to not train and build repository-specific models if there is not enough training data. While this primarily helps us in increasing the model's accuracy and efficiency, it also has the effect of reducing the load on the training and retraining pipeline.
We have implemented Nudge using a map-reduce-based big data platform which will enable us to scale to 1000s of repositories in the future.

\paragraph{Notification presentation}
We experimented with several versions of the notifications. The most verbose explained what features the model looks at, what the estimated lifetime for the \pr{} is, and why we are nudging at a given point in time. A less verbose version just says ``This \pr{} has been open since $N$ days. Please take appropriate action.'' We also experimented with the format, icon, color, etc. We experimented with the different designs of the notification by letting real users try them. Eventually, this helped us come up with a notification that is liked and approved by the end-users.

\paragraph{Feedback collection}
In order to enable ourselves and repository owners to monitor and evaluate Nudge's usage and impact,
we include a feedback collection mechanism.
We rely on thumbs up/down feedback as well as optional text left by pull request authors and reviewers.
A collection pipeline scrapes this feedback automatically per repository.
We also built an internal reporting tool with a dashboard that displays the feedback at the repository level as well as globally, and which is refreshed automatically when the numbers are updated.

\section{Pull Request Lifetime Prediction}
\label{sec:prediction}
To be able to `nudge' developers on overdue \pr{}s, the Nudge algorithm, first of all, needs to determine the expected lifetime of the \pr{}.
In this section, we explain the details of how the data needed to train the lifetime prediction models at \pr{} level is mined and how the model is developed, validated, and deployed. We can broadly classify this activity into the following three steps:
\begin{enumerate}
    \item 
    Leveraging the rich history of prior work done in effort estimation software repository mining to determine the factors that impact \pr{} acceptance and defect prediction (see \cite{Boehm:1984:SEE:2090.2283139, Briand:1999:ACC:302405.302647, Chulani:1999:BAE:322677.322688, Yu:2015:WDP:2820518.2820564} as discussed in Section~\ref{RelatedWork}), we identify a set of attributes that needs to be mined for \pr{}s.
    \item 
    We collect data for these selected attributes on multiple repositories, as well as the actual pull request lifetime data, to establish a training data set.
    \item 
    We use the training data set collected in Step 2 to build a pull request lifetime prediction model and evaluate the performance of the model.
\end{enumerate}


\subsection{Correlation Analysis} \label{CorrelationAnalysis}
We performed correlation analysis to understand the factors that are associated with the lifetime of \pr{}s and the magnitude of the association. We collected 22,875 completed pull requests from 10 different repositories at Microsoft. These repositories host the source code of various medium and large-scale services with hundreds to thousands of developers working in those repositories. 
We omit any \pr{}s whose age is less than 24 hours (short-lived \pr{}s) or more than 336 hours (two weeks, long-lived \pr{}s). 
The reason for omitting short-lived pull requests is that they do not need to be nudged. 
We omit long-lived pull requests as they are outliers, and we do not want the model to learn from poorly handled pull requests that took too long to complete.

We formulated this as a regression problem where we define a dependent variable (\pr{} lifetime) and a set of independent variables (the 28 features listed in Table \ref{table-features}). We then used a gradient boosting regression algorithm to perform the regression analysis and calculate coefficients (listed in Table \ref{table-features}).
The dependent variable in our experiment is the \pr{} completion time, i.e., the time interval between pull request creation and closing date, in hours.
We exclude the 48 weekend hours from the total completion time to make the experiment reflect the real-world deployment scenario where \PaperTitle\ notifications are not sent on weekends.
The features we use in our experiment are related to
the \pr{} itself, the author, the process and churn. Table \ref{table-features} lists all the features we use including their correlation to \pr{} completion time.

\begin{table}
\renewcommand*{\arraystretch}{1.2}
 \begin{tabular}{p{9cm}|l|r}

 \textbf{Feature Description} & \textbf{Type} & \textbf{Corr.} \\
 \toprule
The day of the week when the \pr{} was created & Categorical & 0.163 \\
The average time for \pr{} completion by the developer who initiated it & Continuous & 0.159 \\ 
Total number of required reviewers on the current \pr{} & Discrete & 0.131 \\
Is .csproj file being modified? & Categorical (Binary) & 0.103 \\ 
The average time for completion for the \pr{}s which have the same project paths changed & Continuous & 0.089 \\ 
Total number of distinct file types that are being modified & Discrete & 0.084 \\
The word count of the textual description of the \pr{} & Discrete & 0.072  \\ 
Is the \pr{} modifying any config. files or settings & Categorical (Binary) & 0.059 \\ 
Number of active \pr{}s in the repository & Discrete & 0.058 \\
Churned LOC per class & Discrete & 0.055 \\ 
Total churn in the \pr{} & Discrete  & 0.039 \\ 
Number of methods being churned & Discrete & 0.037 \\ 
Is this \pr{} introducing a new feature? & Categorical (Binary) & 0.033 \\ 
Number of lines changed & Discrete & 0.031 \\ 
Number of distinct paths that are being touched in the current change & Discrete & 0.031 \\ 
Number of conditional statements being touched & Discrete & 0.029  \\
Number of loops being touched & Discrete & 0.028 \\ 
Number of classes being added/modified/deleted & Discrete & 0.021 \\ 
Is the PR doing any refactoring of existing code? & Categorical (Binary) & 0.021  \\ 
Number of references or dependencies (on other libraries / projects) being changed & Discrete  & 0.017 \\ 
Number of files that are being modified in \pr{} & Discrete & 0.016 \\ 
Is the \pr{} making any merge changes like forward or reverse integration (FIs / RIs) & Categorical (Binary) & 0.008 \\ 
Is the \pr{} deprecating any old code? & Categorical (Binary)  & -0.001  \\ 
The word count of the textual title of the \pr{} & Discrete & -0.001 \\ 
Whether the \pr{} is created during business hours or off hours? & Categorical (Binary)  & -0.019 \\ 
Is the \pr{} fixing bugs? & Categorical (Binary) & -0.028 \\ 
Time spent by the developer in the current team. & Continuous  & -0.031 \\ 
Time since the first activity in the repository by the \pr{} author & Continuous & -0.046 \\ 
Time spent by the developer at Microsoft & Continuous & -0.056 \\ 
\bottomrule

\end{tabular}
  \caption{\label{table-features}Feature description and the correlation between features and the \pr{} lifetime (sorted in the descending order of correlation)}
\end{table}

Out of the 28 features, the four that contribute most to a pull request’s lifetime include:
\begin{description}
\item[Day of the week] This is the day of the week on which the \pr{} is created. We represent Sunday with 0 and Saturday with 6. A strong positive correlation with this metric indicates that \pr{} created later in the week are taking more time to complete.
Pull requests created toward the end of the week stay idle during the weekend, but, optimistically, reviewers will start to act on them on Monday. 
We represent days towards the end of the workweek with higher values and check if this affects completion time. 

\item[Average duration of \pr{}s created by the author] This captures how quickly a specific author's \pr{}s were moving, historically. Developers  new to a particular repository or project may take more time to learn the processes followed in the repository. Their changes might be subjected to more thorough reviews and testing which potentially delays the progression of their \pr{}s. Over time these developers may become faster in completing their \pr{}s.

\item [Number of reviewers of the \pr{}] If more people are actively reviewing a \pr{} and are engaged with it, more comments and questions are raised. Some teams in Microsoft have policies that mandate the comments to be closed before completing \pr{}s. So, the \pr{} author has to go through the review comments manually and either agree and resolve them or disagree with them.  

\item [Is a .csproj file being edited] A .csproj file in C\# is a crucial project configuration file that tracks files in the current project, external package dependencies and their versions, dependencies among different projects, etc. Modifications to these files tend to indicate a major activity or structural change in the project. That includes adding or deleting files, modifying external dependencies or libraries, bumping up the versions of the dependent libraries or packages, etc.
\end{description}

The four features that help most reduce the lifetime of a \pr{} include:
\begin{description}
\item[Is the pull request a bug fix?] In large-scale cloud service development environments at Microsoft, fixing bugs is prioritized. Incident management processes help in expediting such bug fixes which result in faster completion of \pr{}s. We used the models developed by Wang et al \cite{wang2019leveraging} to determine the intent of the \pr{}s. These are language models which analyze the pull request title and description to classify the intent. We used these Random Forest models to compute this intent feature along with other features (whether the \pr{} is deprecating old code, whether the \pr{} is performing refactoring). This helps account for the semantic intent of the pull request in the lifetime prediction model.
\item[Age of the author in the team] This feature captures how familiar a developer is with the current team, its processes, people, and the product or service the team is working on. The more time a developer spends in a team, the less difficulty they will experience in pushing their change through. We get this information from the human-resources database at Microsoft.
\item[Age of the author in the repository] This helps capture the familiarity of a developer with the repository in which they are making changes, and the build, and deployment processes of that repository. Although this may sound similar to the author's age in the current \emph{team} just discussed,
familiarity with repositories may vary substantially in heterogeneous teams that work on multiple services (especially, microservices).
Here, different members of the same team are mostly making changes that are very specific to the repositories they are actively engaged in. Our correlation analysis has shown that the more familiar a developer is with a specific repository, the less time it takes them to merge their changes made in that repository. We compute this based on when the author created or reviewed the first pull request in a given repository.
\item[Age of the author in Microsoft] This helps capture the seniority of a developer. Intuitively, senior people who have more experience tend to make fewer mistakes and will experience less pushback on their changes. The negative correlation here indicates that if someone has more experience, it takes them less time to merge their changes. We get this information from the human resources database at Microsoft.
\end{description}

\subsection{Prediction Model} \label{PredictionModel}

\begin{table}
\setlength\belowcaptionskip{7pt}
\caption{Comparison of different prediction models}
 \begin{tabular}{l|r|r} 
 \toprule
Algorithm & MAE (in hours) & MMRE \\ 
\midrule
Least squares & 44.32 & 0.68 \\ 
Bayesian ridge & 46.35 & 0.71 \\ 
Gradient boosting & 32.59 & 0.58 \\ 
\bottomrule
\end{tabular}
\label{model-comparison}
\end{table}

As indicated, we approach the task of predicting the lifetime of a \pr{} as a regression problem. 
We  include most of the features from Table~\ref{table-features}, dropping the ones with a very low (absolute) magnitude of correlation. We used 0.008 as a cut-off, thus dropping three features. This helped to speed up the training and inference tasks without materially impacting the MAE (it dropped by 10~minutes (0.17~hours)).

We then performed an offline analysis and evaluation with multiple popular regression algorithms like least-squares linear regression, Bayesian ridge regression, and gradient boosting. To compare the regression algorithms, we used two standard metrics: MAE (Mean Absolute Error), and MMRE (Mean Magnitude of Relative Error). These metrics are widely used for understanding the performance of regression tasks. We decided to adopt gradient boosting as it has better accuracy with respect to both MAE and MMRE. The comparative analysis of the three algorithms, evaluated against MAE and MMRE is shown in Table~\ref{model-comparison}. A detailed discussion on prediction accuracy and its significance in the context of the application we are building is presented in Section \ref{sec:eval}.

We are not using the prediction outcome (the expected \pr{} lifetime) for performing traditional effort estimation tasks, such as sprint or project planning or budgeting.
In the case of the Nudge system, the primary purpose of the model is to approximate the opportune moment to send a reminder. 
Therefore, the Nudge system exhibits more tolerance toward the prediction error.

To make sure all the features reflect recent trends, we use the \pr{} data from the past two years each time the model is trained.

For training and evaluation, we use scikit-learn\footnote{\url{https://scikit-learn.org/}}.
We used a standard 10-fold cross-validation. We followed the standard practice of one time 10-fold cross-validation \cite{CrossValidation}, without `repeated cross-validation', as follows:
\begin{enumerate}
    \item We separate the data set into 10 partitions randomly;
    \item We use one partition as the test data and the other nine partitions as the training data;
    \item We repeat Step-2 with a different partition than the test data until all data have a prediction result;
    \item We compute the evaluation results through a comparison between the predicted values and the actual values of the data. 
\end{enumerate}

\section{\pr{} status determination} 
\label{sec:status}

With a mechanism in place to predict the lifetime of a \pr{}, the next step is assessing whether there has been any activity or state changes that are taking place in a \pr{}. 
This serves to determine the opportune moment to send a notification as well as to understand when \emph{not} to send a notification.
To do so, we determine the current \emph{activity} and \emph{blocking actor} 
in terms of the pull request lifecycle model as displayed in Figure~\ref{PrStateDiagram},

\subsection{Activity detection} \label{ActivityDetection}

Using an earlier version of Nudge, we conducted a quantitative study to understand the impact of not reacting to the activity in a \pr{} while sending notifications. We found, through manual inspection, that 86 out of 119 Nudge comments that are resolved negatively were due to the fact that Nudge did not honor the recent \pr{} activity. Later, we talked to some of the developers who were either authoring or reviewing those \pr{}s. 
The majority of them did not like the Nudge notifications because they recently interacted with the \pr{}. 

To resolve this problem, Nudge determines the most recent activity in the \pr{}s. However, \pr{}s in large organizations can get complex with multiple actors performing different activities through various collaboration points. We distinguish the following collaboration points that trigger the changes to the \pr{}s (see also Figure~\ref{PrStateDiagram}):

\begin{description}
\item[\prc{} state changes] A state change in a \pr{} strongly indicates that one of the actors (author or reviewer) has been acted on the \pr{} recently.

\item[Comments] Once a \pr{} is submitted for review, reviewers can add comments to recommend changes or seek clarification on a specific code change. Authors of the \pr{} can also reply to the comment thread that is started by the reviewers if they have any follow-up questions. In addition to placing the comments and replying to them, the actors can also change the status of the comments. Typical statuses are `Active' which means the comment has just been placed, `Resolved' which means the comment has been resolved by the author of the \pr{} by making the changes prescribed by the reviewers, `Won't fix' which means the author would like to discard the review recommendation without addressing it, and, `Closed' which means the comment thread is going to be closed as there are no more follow up action items or discussions needed.

\item[Updates] After a \pr{} has been created, authors can keep pushing new updates in the form of commits. These commits are changes that authors are making in response to review recommendations or improvements the authors themselves decided to push into the \pr{}. Under some special circumstances, someone other than the author or the reviewer can also push new updates into a \pr{} but that is a rare occurrence. New updates or iterations are a very strong indicator that the author is making progress on the \pr{}.

\end{description}

The specific action points may vary depending on the provider of the source control system (GitHub, Azure DevOps, GitLab, etc). However, conceptually the collaboration points or concepts remain similar. In the context of this work, we focus on Azure DevOps, the source control system used by Microsoft's developers and offered by Microsoft to third-party customers. We track the activities performed through these collaboration points to determine the existence of activity in \pr{}s and decide whether a Nudge notification should be sent.

Nudge typically sends a notification once the lifetime of a \pr{} crosses the predicted lifetime (as predicted by the lifetime prediction model). However, it waits at least 24 hours before sending a notification, if there has been any activity since last checked (Nudge pipeline runs once every six hours; details about the pipeline are explained in Section \ref{Implementation}) or when state transitions have been observed. Based on a user study described in Section \ref{sec:eval}, we find that activity detection improves user experience, reduces false alarms, and thus increases the usefulness of the Nudge service.

\subsection{Actor Identification} \label{DependecyDetermination}

In a \pr{}, there are different actors involved (as explained in Section \ref{Background}) that can influence the next state (Approved, Rejected, etc.),
and the speed with which a \pr{} progresses.

\begin{table}
\centering
\setlength\belowcaptionskip{7pt}
\caption{Classes that explains change blockers and responsible actors}
\begin{tabular}{p{1.5cm}|p{7cm}|p{1.75cm}|r} 
\toprule
State & Class & Actor \par waiting for & \#PRs \\
\midrule
Waiting & Not all review comments have been addressed & Author & 34 \\
Waiting & Pull request needs further discussion & Author & 47 \\
Approved & Pull request has been approved but the author is not ready to merge it & Author & 49 \\
Active & Review has not been started yet & Reviewer & 51 \\
Active & Review comments have been addressed but reviewers have not approved yet & Reviewer & 19 \\
\bottomrule
\end{tabular}
\label{PRBLockers}
\end{table}

\begin{figure}
\centering
\includegraphics[width=0.95\columnwidth]{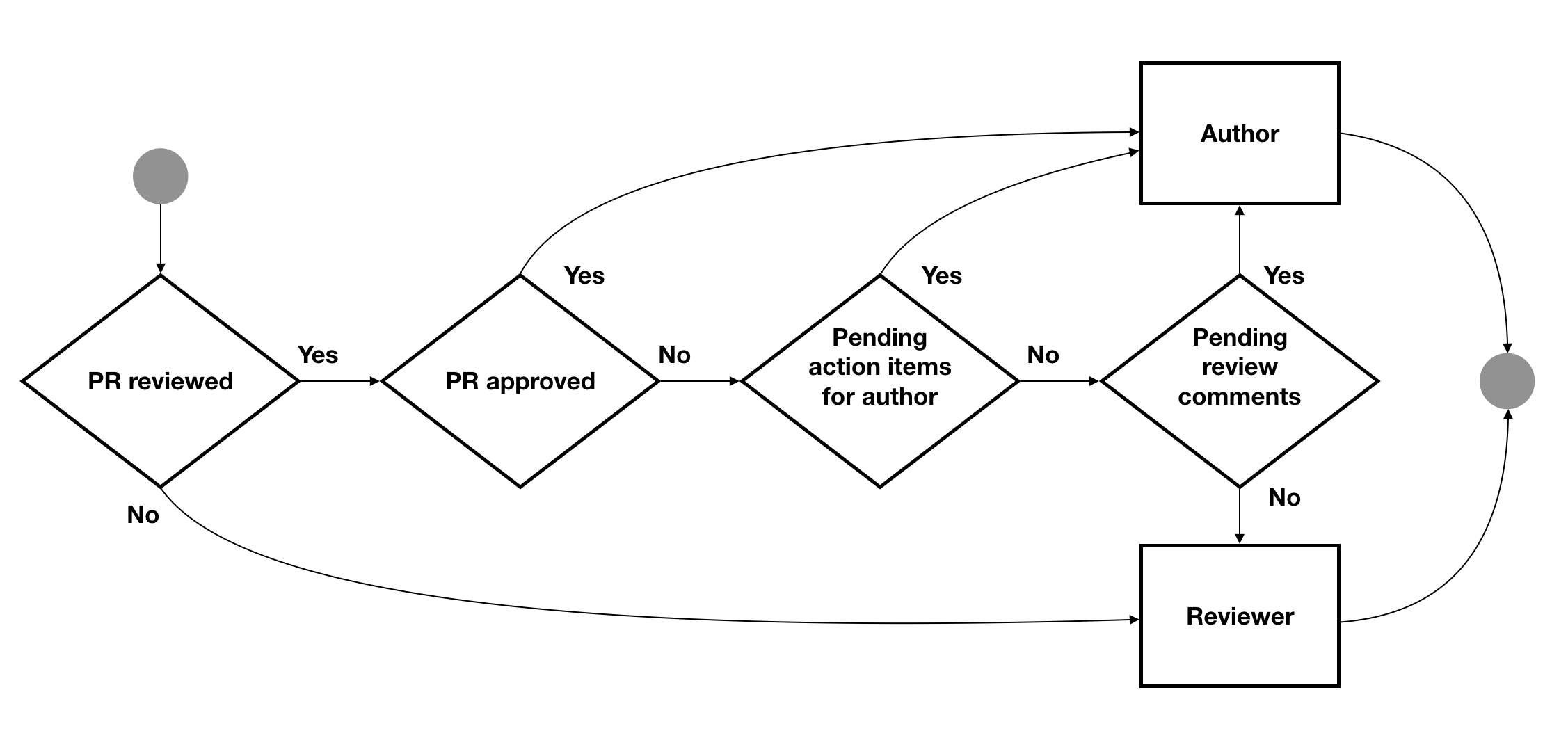}
\vspace{-0.5\baselineskip}
\caption{Flowchart to determine the change blockers for active \pr{}s}
\label{fig:Dependency Flowchart}
\end{figure}

We focus on understanding the human change blockers, i.e., authors and reviewers, and the extent to which they influence the change progression.
We collected 200 \pr{}s from 20 medium to large to very large repositories and manually analyzed them to understand for whom they were waiting before they were completed.
These are \pr{}s whose age is at least 14 days and which have not been completed yet. We find there are five mutually exclusive classes that explain the cases in which a \pr{} is awaiting completion. 
Table~\ref{PRBLockers} lists the classes and the actor responsible, and the number of \pr{}s that fall under each class. 70~\pr{}s (out of~200) are blocked by the reviewers while the remaining 130 are waiting for the author to make progress.

Encouraged by the findings, we devised an algorithm that helps determine the actor that needs to be notified to make progress on a given \pr{}. When there is an action item pending on the author of the \pr{} as well as a reviewer, sending notifications to the author is prioritized. The flow chart shown in Figure \ref{fig:Dependency Flowchart} explains the control flow and how the actors that are responsible for making progress on the change are determined.
The algorithm evaluates various decision points to determine the blockers of a change. These decision points represent different states that a \pr{}, review, or reviewing comments in the \pr{} take during the lifetime of a \pr{}.
There are three cases where the author needs to act:

\begin{description}

\item[PR is approved] A \pr{} is approved when the reviewers are satisfied with the changes and have no more comments or concerns about the change. The author can proceed to merge the change.

\item[Not all review comments are addressed] The reviewer has left comments seeking some clarity or proposing recommendations. The author is responsible to address the review comments. Authors typically will have two choices: If they agree with the review comment, they can resolve it or if they disagree, they can mark it as `won't fix'. This condition is met if the author has review comments that need to be addressed.

\item[Author has pending action items] The author has addressed the review comments but the reviewer does not want to approve the changes, because they are not satisfied with the resolution provided by the author. These \pr{}s need further discussion.
\end{description}

In the remaining cases, the reviewers have to act on the \pr{} to unblock the change:

\begin{description}
\item[Review has not started] Upon creating the \pr{}, authors typically add the reviewers that they would like to get a review from for the specific change. The reviewers are supposed to act on it and provide their comments. If the reviewers are not acting on the \pr{} after requesting a review the onus is going to be on the reviewers to act on the \pr{} and unblock it.

\item[Review comments are addressed] Once the reviewer has provided their review, the author will act on it and resolve/won't fix the comments by making necessary changes. Then the responsibility shifts back to the reviewer to re-verify the changes and sign off the change. If that is not happening, reviewers are accountable and should be notified to unblock the change.

\end{description}

\section{Implementation} \label{Implementation}
In this section, we present the details about how the \PaperTitle\ Service is implemented. 
It relies on \ado{}, the git-based DevOps solution offered by Microsoft, which we used to deploy \PaperTitle\ as an extension. 

\subsection{\PaperTitle{} Service Architecture}

Figure \ref{fig:architecture} shows the \PaperTitle{} service architecture and gives an overview of various components involved. 
\ado~ is the existing git environment, which is connected to a collection of workers hosted on Azure.
Listed below are the seven steps (the numbered arrows in the figure) that explain the high-level architecture and interaction between various components in the \PaperTitle\ system:

\begin{figure}
\includegraphics[width=1\columnwidth]{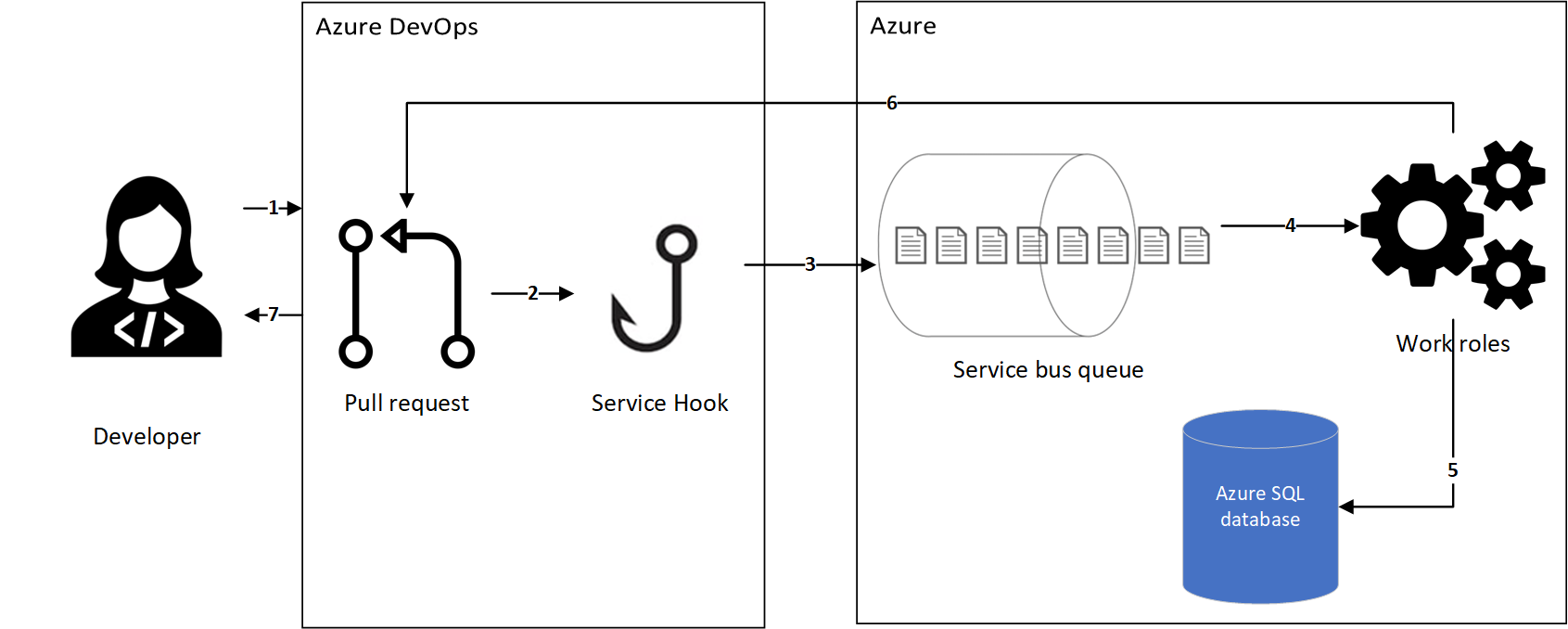}
\caption{\PaperTitle\ Architecture}
\label{fig:architecture}
\end{figure}

\begin{enumerate}
    \item A developer creates a \pr{} or updates an existing \pr{} by pushing a new commit or iteration into it.
    \item \label{step:update} A \pr{} creation or update event is triggered through the service hook.
    \item A new message is sent to the Azure service bus where it is queued.
    \item An Azure worker picks up the new messages on a first come first serve basis.
    \item \label{step:worker} The workers run effort estimation, activity detection, and actor identification, storing the results in a database.
    \item Based on the outcome of step~\ref{step:worker}, if the Nudge system decides to send a notification, the worker sends a notification using the \ado~ APIs in the form of \pr{} comments.
    \item \ado~ sends a notification email to the developer.
\end{enumerate}

Nudge performs inference and recalculates the effort each time a \pr{} is updated (Step~\ref{step:update}). PR updates can change the structure of the PR completely (adding/deleting code changes, adding/removing reviewers, etc.), making it important to react to them and adjust the pull request lifetime prediction accordingly.

Additionally, if the criteria to send a notification are not met for a given \pr{}, the Nudge system will check again in six hours, through an Azure batch job, to determine if it can send a notification. The Nudge service continues to do that, every six hours, until the \pr{} is abandoned or completed.

\subsection{Azure DevOps} 

\ado{} is a platform providing a git-based version control system. In addition to repositories, it offers planning tools such as work item and bug report management and facilitates code review management. It also has features such as build and releases management to facilitate continuous integration and deployment. The \PaperTitle\ service is deployed as an extension of \ado{} because of the rich collaboration features offered by \ado{}. Below are the details about some of the key features that \ado{} offers that helped materialize the \PaperTitle\ service: 
\begin{description}
    \item[Collaboration points:] \ado{} offers a rich set of collaboration points through which third-party services or extensions can interact with \pr{}s in \ado{}. The collaboration points allow services to add comments on \pr{}s, add labels to the \pr{}s, and add or remove reviewers.
    
    \item[Service hooks:] \ado{} offers service hooks that help any third party service to listen to the events that are happening inside the \pr{} environment. Events can be \pr{} creation events that are fired through service hooks when a \pr{} is created or \pr{} update events that are fired when the \pr{} experiences any updates such as pushing new commits or iterations.
    
    \item[APIs:]\ado{} exposes a rich set of REST APIs ~\cite{AzureDevOpsRest} that helps third-party services to access information about various artifacts in the \ado{} environment. These APIs can be called through a REST client and return metadata about the \pr{}s (id, title, author, reviewer information, comments, labels, status, commits that are included in the \pr{}), commits (title, files changed in a commit), build and release (status, test outcomes, deployment outcomes).
    
    \item[Votes:] Azure DevOps uses a voting mechanism to capture the actions performed by the reviewers on a \pr{}. A vote on a \pr{} can have values $\{-10, -5, 0, 5, 10\}$, corresponding to
    rejected, waiting for the author, no vote, approved with suggestions, and approved, respectively.
\end{description}

\subsection{Activity detection} We use \ado{}'s REST APIs \cite{AzureDevOpsRest} to collect data that is required to understand if there has been any activity in a \pr{}. We gather data about various actions or activities that happen inside a \pr{} (Section \ref{ActivityDetection}) to determine if there has been any activity, as follows:

\begin{description}
    \item[Commit activity:] We use \ado{}'s \texttt{GetPullRequestIterationsAsync} API which provides details about all the commits that are ever pushed into a \pr{}. We first get a list of all the commits that are pushed and then take the timestamp of the latest iteration as the latest commit timestamp of a \pr{}.
    \item[Comment activity:] To determine whether there has been any commenting activity like adding new comments or replying to existing comments, we use \ado{}'s \texttt{GetThreadsAsync} API. This API returns all the comments that are ever placed in a \pr{} in the form of threads. We check if any new threads are created or if any new comments are placed in an existing thread. We take the maximum of both of them to determine the latest comment activity that has happened in a \pr{}. While doing this we exclude any comments that are placed by system accounts or non-human actors following basic heuristics, such as accounts that include words like 'system', 'bot', 'account', etc.
    \item[State changes in \pr{}s:] Changes in \pr{} state is another important signal that helps determine activity in a given \pr{}. Unfortunately, there is no direct way of determining state changes in \pr{}s. We use \ado{}'s \texttt{GetThreadsAsync} API to collect all the comments placed in a \pr{}. Comments whose content property contains the word ``voted'' indicates that a state change has happened. Azure DevOps uses a voting mechanism to capture the actions performed by the reviewers on a \pr{}. A voting event in a \pr{} looks like the following: ``\texttt{User1 voted 10 on PR1234}',
    which, as explained above, corresponds to approval.  We use such events to determine the last time a \pr{}'s state has changed.
\end{description}

Nudge sends a notification once the lifetime of a \pr{} crosses the predicted lifetime (Section~\ref{sec:prediction}). However, it waits at least 24 hours before sending a notification, if there has been any activity observed.

\subsection{Actor identification} 
We rely on \ado{}'s REST APIs to collect data for identifying the actors.
In line with Section~\ref{DependecyDetermination}, 
we use \ado{} as follows:

\begin{description}
    \item[Check pending action items:] To determine if \pr{}'s author has any pending action items, we check if the state of the \pr{} is set to ``Waiting on Author''. We use \ado{}'s \texttt{GetPullRequestReviewersAsync} API to get the votes of all the reviewers.
    \item[Check for existence of unresolved comments:] The existence of unresolved comments determines whether the blocker of a \pr{} is the author or the reviewer. We use \ado{}'s \texttt{GetThreadsAsync} API to get all the threads. We then check for the existence of threads with statuses ``Active'' or ``Pending''. The presence of threads with any of these two statues indicates that there are unresolved comments.
    \item[Enumerate the list of change blockers:] We first use the \texttt{GetPullRequestReviewersAsync} API offered by \ado{} to query the list of reviewers on a given \pr{}. We then use the \texttt{GetThreadsAsync} API to determine the list of all the reviewers who commented on the \pr{} at least once and whose comments are resolved by the author of the \pr{}. We prepare two lists: reviewers who commented, and all reviewers, and choose one of them to use based on the state of the \pr{}. If there are no reviews on a \pr{}, we send notifications to the reviewers in the `all reviewers' list. If there has been a review activity (reviewers placed comments on the \pr{}), we prioritize notifications to the reviewers in the list of reviewers who commented.
\end{description}

\subsection{Nudge Notification}

\begin{figure}
\includegraphics[width=0.8\columnwidth]{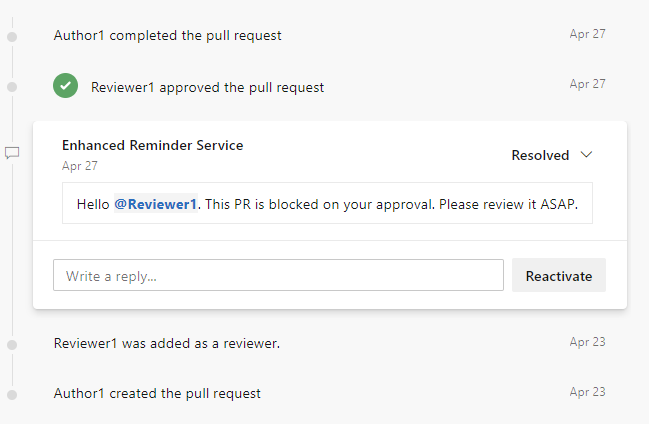}
\caption{\PaperTitle\ notification, with @Reviewer1 tagged in the reminder}
\label{fig:Screenshot}
\end{figure}

Figure \ref{fig:Screenshot} shows the screenshot of the Nudge notification. 
Note that the dependent actor (in this case the reviewer but not the author) is being ``@-mentioned'' in the notification. This triggers a separate email to the reviewer of this \pr{} asking them to unblock the \pr{}. As we can notice, the \pr{} was created and had been waiting for the reviewer's approval for four days. After the \PaperTitle\ service tagged the reviewer and pushed them to act on the \pr{}, the reviewer approved it and the \pr{} got completed on the same day.

\section{Evaluation} \label{sec:RQs} \label{sec:eval}
In this section, we describe 
(1) the experiments we conducted to assess the value of a \pr{} level effort estimation system, 
(2) the value of a system like Nudge that leverages the effort estimation models to notify developers about their overdue \pr{}s, 
(3) the impact Nudge has on large development teams and organizations, and
(4) the scale at which a system like Nudge can operate.
This is reflected in the following research questions:
\begin{description}
    \item[RQ1] What is the accuracy of effort estimation models in predicting the lifetime of \pr{}s?
    \item[RQ2] What is the impact of a service like \PaperTitle{} on completion times of \pr{}s?
    \item[RQ3] What are developers' perceptions about the usefulness of the \PaperTitle\ service?
    \item[RQ4]
    Can the deployment of Nudge be scaled to thousands of repositories
    without sacrificing gains in pull request processing time and user perception?
\end{description}

\subsection{Data collection and methodology}
We obtained data from the large-scale deployment of the Nudge service for nine months on \NumRepos\ repositories in Microsoft. 
The data includes telemetry from the Nudge service using only lifetime prediction as a mechanism (which we will refer to as Nudge-LT), as well as from the Nudge service extended with activity and actor identification (which we refer to as Nudge-FULL or just Nudge).
The repositories are owned by various product and service teams and are of different sizes, geographies, and products. 
Nudge has made notifications on \NumNudgedPrs\ \pr{}s during the nine-month time window under study. 
We discuss the results from Nudge-LT first and subsequently analyze the effect of the additional heuristics of Nudge-FULL.


For RQ1, we collect historical data from \pr{}s that are merged. This gives us the start and end timestamps to help us calculate the lifetime for each \pr{} and construct a ground truth data set. We collected 2,875 pull requests from 10 different repositories that have been merged and completed. These repositories host the source code of various services. Their number of contributing developers ranges from a few hundred to a few thousand.
This data set is independent of the data used to train the model (as explained in Section \ref{PredictionModel}). 
As for the correlation analysis (Section~\ref{CorrelationAnalysis}), 
we omit any \pr{}s whose age is less than 24 hours (short-lived \pr{}s) and more than 336 hours (long-lived \pr{}s).

\begin{figure*}
\includegraphics[width=\textwidth]{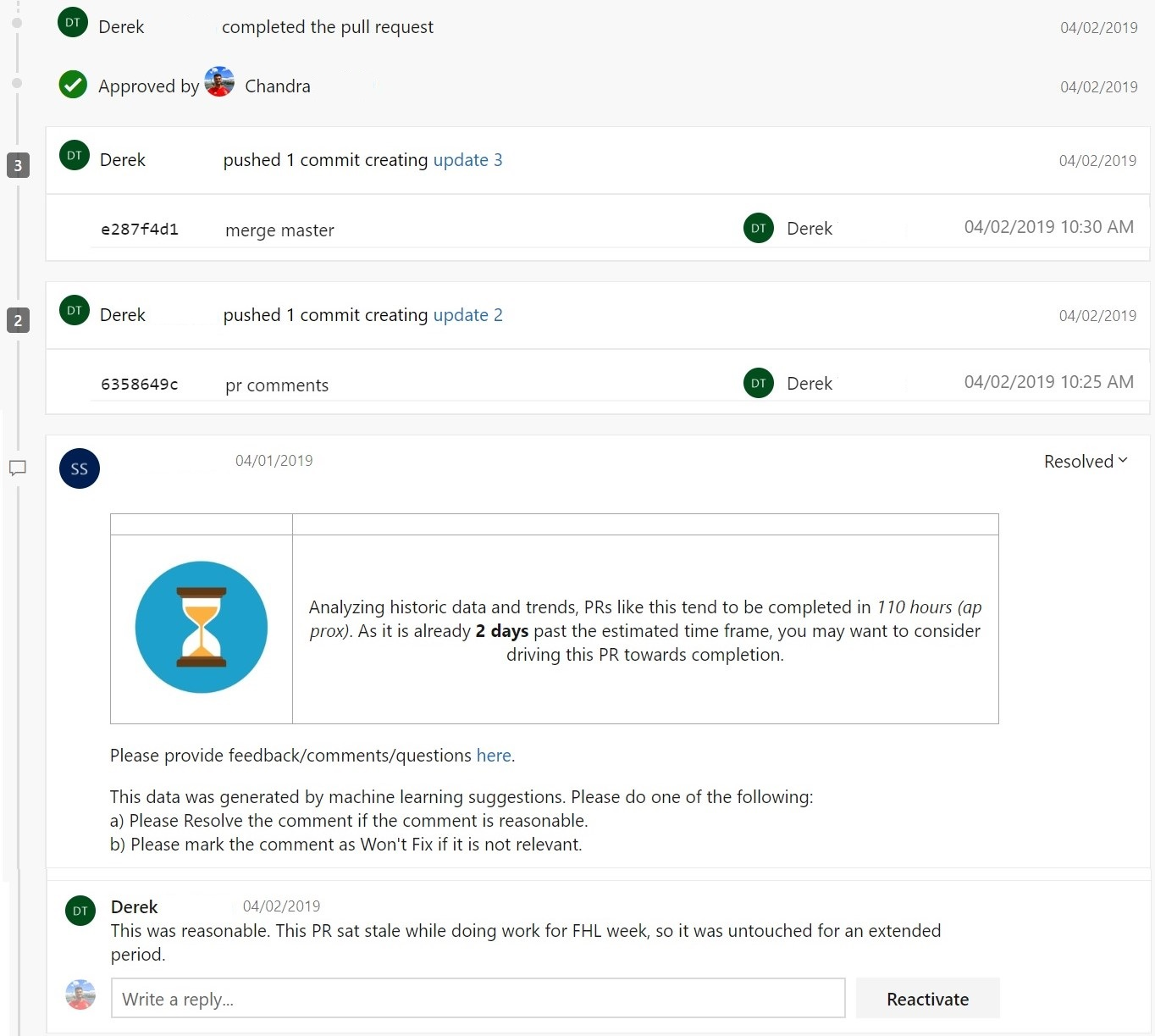}
\caption{A pull request with a lifetime prediction notification in Azure DevOps}
\label{fig:screenshot}
\end{figure*}

For RQ1, we also use repositories on which we operationalized Nudge to obtain feedback from developers on the estimations.
We randomly select \pr{}s for which we are about to send a Nudge notification and add more details in the notification comment. These are details like Nudge model's predicted lifetime for a given \pr{}, how long the \pr{} has been open past the estimated lifetime by the Nudge model. Figure \ref{fig:screenshot} shows the details about the predicted lifetime of a sample \pr{}, as predicted by the Nudge model, and the reason for sending the notification at a given point in time. 

For RQ2, we collect data from the \NumRepos\ repositories on which we operationalized Nudge. We collect data on how the lifetime of \pr{}s is varying between \pr{}s which received a Nudge notification and \pr{}s that did not. We also collect data about the time it takes for the author of the \pr{} to either complete or abandon the \pr{} after a Nudge notification is sent.

For RQ3, we collect data through our automated pipeline that actively tracks every single inline reply that is posted by the developers in response to a Nudge notification and whether they positively or negatively resolved a comment. We do this for all \NumNudgedPrs\ \pr{}s on which we made notifications.

For the \NumRepos{} repositories on which we deploy Nudge, 
notifications are sent when a \pr{} meets the criteria needed to be nudged, as imposed by the Nudge model and algorithm.
All developers who receive a Nudge notification are given equal opportunity to provide feedback in the pull request, either to positively or negatively resolve the comment or to provide anecdotal feedback by replying inline to the Nudge notification. 
Note that the repositories on which Nudge has been operationalized are organizationally away from the developers of the Nudge service. The notifications did not reveal the names or identities of the developers of the Nudge service to avoid response bias \cite{Cutrell-Bias}.

For RQ4, we took advantage of the fact that our initial experiments convinced Microsoft management to deploy Nudge in production.
This enabled us to monitor Nudge in production at Microsoft during the period January 2021 until December 2021. During this period, Nudge was deployed on 8,000 different systems at Microsoft. Nudge sent 210,000 notifications authored by 40,000 unique developers.
We collect pull request completion time after notification, as well as the positive/negative resolutions of pull request recommendations made by Nudge.

\subsection{RQ1: What is the accuracy of effort estimation models in predicting the lifetime of \pr{}s? } \label{RQ1}
To answer this research question, we collect metrics that explain how accurate our prediction model is. 
We also list the anecdotes we received from the developers about the accuracy of the prediction model. 

\begin{figure}
\includegraphics[width=0.8\hsize]{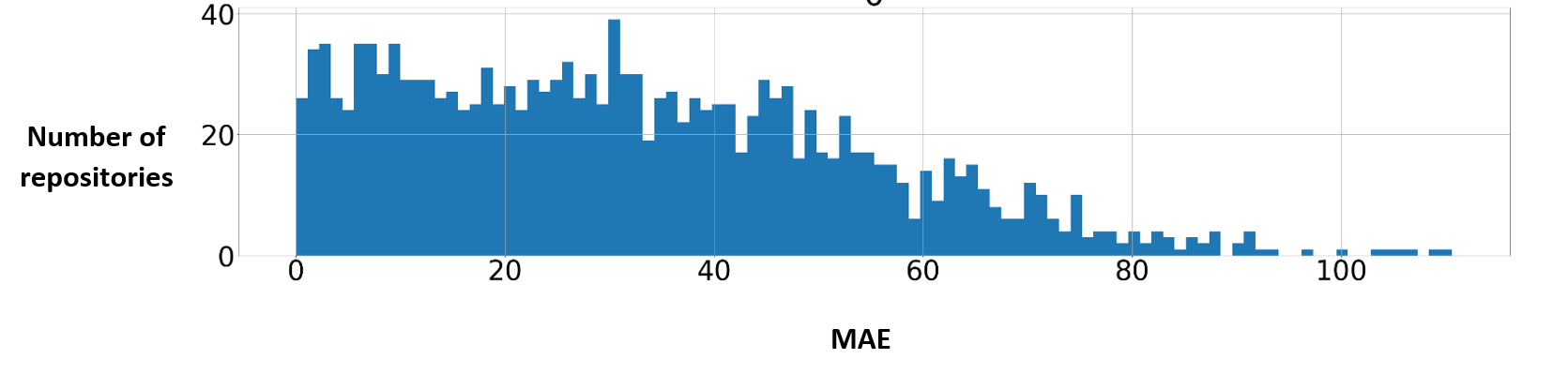}
\caption{MAE distribution}
\label{fig:mae}
\end{figure}
\begin{figure}[b]
\includegraphics[width=0.8\hsize]{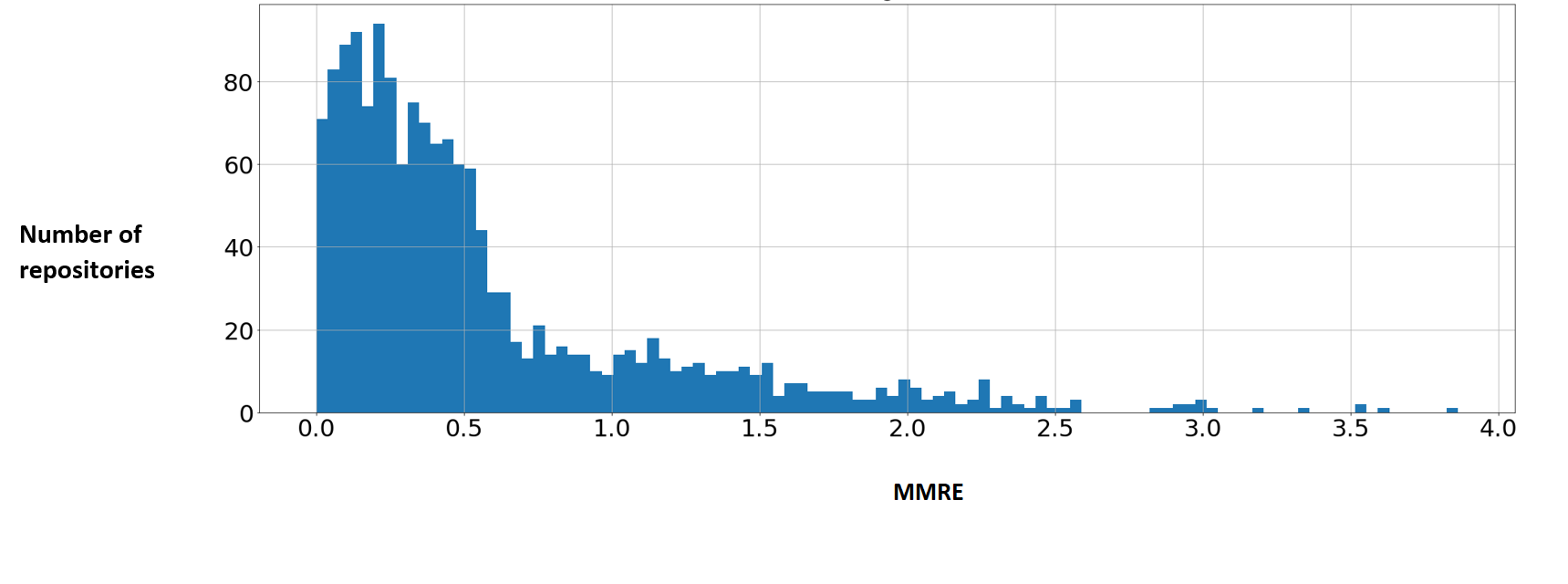}
\caption{MMRE distribution}
\label{fig:mre}
\end{figure}

\paragraph{Model Evaluation}
 We evaluated our prediction model against standard metrics: Mean Absolute Error (MAE) and Mean Magnitude of Relative Error (MMRE). For the \pr{} level effort estimation model, the MAE is 32.60 hours (Figure~\ref{fig:mae} shows the distribution of MAE) and MMRE is 0.58 ( Figure~\ref{fig:mre} shows the distribution of MMRE). 
 
 To put these numbers in perspective, we have conducted an experiment by considering the mean lifetime of our training data as the predicted lifetime of every \pr{} in our testing data. Our constant model's MAE is 36.43 hours and MMRE is 0.68. This means our trained model is 11.8\% better in terms of MAE and 17.7\% better in terms of MMRE compared to the constant model. 

The MAE of 32.60 hours corresponds to around 1.3 days. The average duration is 107.63 hours or a little over 4 days. For our purposes, for warning developers when they are late, we consider an average deviation of around a day to be acceptable.


\paragraph{User feedback about model's prediction accuracy}
We received positive feedback from the developers of the randomly selected \pr{}s for which we added more details about the model prediction as illustrated in Figure \ref{fig:screenshot}. One of the developers said:
\begin{quote}
   \textit{This was reasonable. This \pr{} sat stale while doing work for FHL, so it was untouched for an extended period.}
\end{quote}
Here, the developer is acknowledging that the model's prediction (110 hours) is reasonable and, noticeably, provides an explanation for why the \pr{} is taking long to wait. As we see in the figure\ref{fig:screenshot}, the developer ended up completing the \pr{} within a few hours after the notification was sent. Similarly, another developer said:

\begin{quote}
   \textit{I totally agree with the model saying this \pr{} should take not more than 120 hours to complete. The code change is slightly complex and the estimation seems reasonable.}
\end{quote}

In this case, the developer is positive about the fact that the model is predicting the lifetime by taking into account the complexity of the change and giving enough breathing room for the developers to act on it before nudging them. Another developer passed feedback by acknowledging the fact that the model adapts to the changes happening inside the \pr{} by comparing two of her \pr{}s.

\begin{quote}
    \textit{I see the estimation is 176 hours on this \pr{} and it was 64 hours on another \pr{} of mine where I was editing a lesser number of files and not pushing critical code changes. I do not know if your model is taking these facts into account. But, it seems like...interesting!!}
\end{quote}

This anecdote supports the fact that the model adapts to the \pr{} in question and the users starts to notice that the model is doing a reasonable job in adapting to the change in context.

\subsection{RQ2: What is the impact of \PaperTitle\ service on completion times of \pr{}s?}
To measure the impact of Nudge, we use two metrics to assess whether the Nudge service is helping developers and yielding the intended benefit:
\begin{enumerate}
    \item Average \pr{} lifetime: This is the average of the time difference (in hours) between \pr{} creation and closing date. A service like Nudge is expected to introduce positive effects like reduction in \pr{} lifetime by notifying the change blockers about making progress and closing the \pr{}s.
    
    \item Distribution of the number of \pr{}s that are completed within a day, in three days, within a week, and after a week since Nudge sent a notification. This captures to what extent developers are  actually reacting to the Nudge notifications, and if so, how quickly they are reacting.
\end{enumerate} 

While measuring and comparing the metrics above, we make sure to nullify the effects of other variables such as the month of the year (changes move faster in some months and slower during some), typical code velocity in a given repository (some repositories naturally experience higher development velocity because of the nature and critically of the service), team or organization culture (some teams typically are more agile and ship things faster), etc. 
Therefore, if we compare \pr{}s from two different repositories or from two different time periods, we cannot confidently say whether an increase or decrease in average lifetime is due to the presence or absence of the Nudge service or due to other factors explained above. 
To remedy this, we set up a randomized trial (A/B, or in fact A/B/C testing) by randomly selecting one of the three configurations listed below for each \pr{}:

\begin{description}
    \item[None:] Turn the Nudge service off for a set of randomly selected \pr{}s.
    \item[Nudge-LT:] Turn on the basic version of the Nudge service with just lifetime prediction, but without user identification and activity detection.
    \item[Nudge-FULL:] Turn on the user identification and activity detection features along with the effort estimation model in the Nudge service.
\end{description}

\begin{table}
\caption{Comparison of average \pr{} lifetime (hours)}
\centering
 \begin{tabular}{l|r|r}
\toprule
Service & Avg PR lifetime & Number of PRs \\ 
\midrule
None & 197.2 & 3856\\ 
Nudge-LT & 112.6 & 4117 \\ 
Nudge-FULL & 77.7 & 4383\\ 
\bottomrule
\end{tabular}
\label{AverageLifetime-comparison}
\end{table}

Table \ref{AverageLifetime-comparison} displays the average \pr{} lifetime for each of these configurations.
We see a clear decrease in an average lifetime for the \pr{}s for which Nudge notifications are sent. The average lifetime of the \pr{}s on which Nudge notifications are sent is 112.6 hours which is a 42.9\% decrease compared to the set of \pr{}s on which we did not send the notification (where the average lifetime is 197.2 hours). Actor identification and activity detection further brought the average lifetime down to 77.7 hours which is a reduction of 60.62\% in average \pr{} lifetime.

\begin{figure}
\includegraphics[width=0.75\columnwidth]{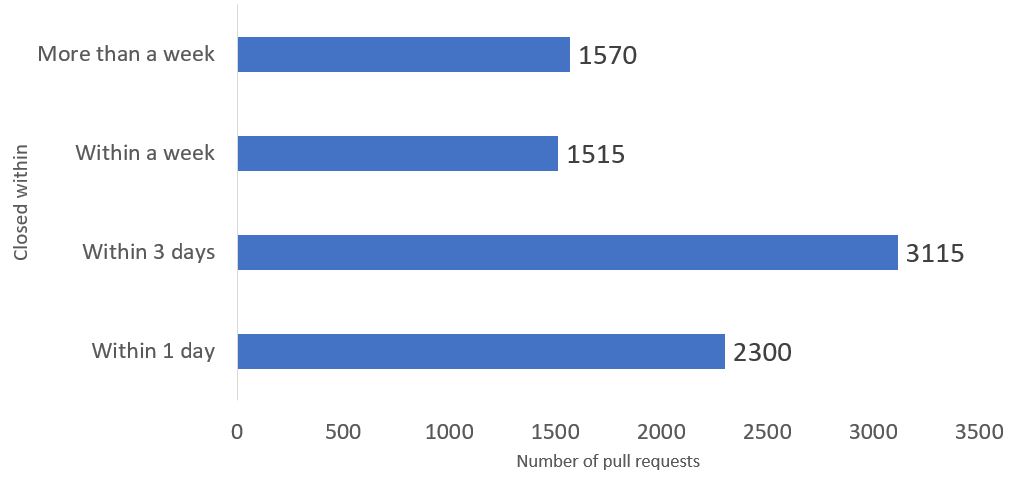}
\caption{Distribution of completed \pr{}s after sending a notification}
\label{fig:prcompletion-times}
\end{figure}

In figure \ref{fig:prcompletion-times}, we plot the distribution of \pr{}s that are completed within a working day, three days, a week or more than a week after Nudge sent the notification. Only 1570 \pr{}s out of 8500 \pr{}s (18.47\%) have taken more than a week to close. 81.53\% of the \pr{}s are closed within a week. An important observation to make is that 2300 \pr{}s i.e., 27.05\% of the \pr{}s on which Nudge sent the notification were completed within a day. This distribution indicates that the majority of the \pr{}s on which Nudge sends notifications are completed relatively quickly.

\subsection{RQ3: What are developers' perceptions about the usefulness of the \PaperTitle\ service?}
To understand whether or not users are favorable towards the Nudge system, we pursue a mixed-methods approach. 
To that end, we rely on two sources of information:

\begin{itemize}

    \item For every Nudge notification that is sent, the developers have an option to perform one of the following three actions: positively resolve the notification (by marking it as `resolved'), negatively resolve the notification (by marking it as `won't fix'), and provide no response.
    \item Second, Nudge users can enter an inline reply within a Nudge notification, to explain their (dis)satisfaction.
\end{itemize}

We again distinguish between Nudge-LT and Nudge-FULL.


\subsubsection{Notification Resolution}
\label{sec:resolution}
Table \ref{Comments-comparison} shows the number of positive and negative reactions to notifications, both for Nudge-LT and Nudge-FULL.
For the vast majority (93\%-97\%) the developers actually provided an explicit verdict.

For Nudge-LT, the majority of verdicts (2062/3891, 53\%) were negative.
This suggests that nudges based on lifetime predictions alone are not considered sufficiently helpful.

For Nudge-FULL, by contrast, the vast majority of verdicts (3199/4018, 80\%) were positive.
When also including non-responses in the total, the percentage of positive resolutions remains high, at 73\% (3199/4383).
This makes it clear that the activity detection and actor identification of Nudge-FULL clearly contribute to the positive perception of Nudge.

%
%
Note also that positive feedback of 73\% is substantial if we look at it in isolation. 
Various studies have shown that users tend to provide explicit negative feedback when they do not like or agree with a recommendation while not so explicit about positive feedback \cite{10.1145/2043932.2043957, 10.1145/3289600.3291003}. 73\% of the developers who received Nudge notifications explicitly resolving the notifications positively indicates a clear positive sentiment that the developers exhibit towards the Nudge service.

\begin{table}
\caption{The difference in percentage of positively resolved notification}
\centering
 \begin{tabular}{l|r|r|r|r|r} 
 \toprule
Service type 
  & \thead{\# Positive\\ responses} & \thead{\# Negative\\ responses} & \thead{\# Total\\ responses} & \thead{\# No\\ responses} &  \thead{\# Total\\ PRs}\\ 
\midrule
Nudge-LT 
  & 1829 & 2062 & 3891 & 226 & 4117 \\ 
Nudge-FULL 
  & 3199 & 882 & 4081 & 302 & 4383 \\ 
\bottomrule
\end{tabular}
\label{Comments-comparison}
\end{table}

\subsubsection{Nudge-LT user feedback} \label{QualitativeFeedback}
We tried to understand how helpful our suggestions are and whether they are yielding intended benefits, i.e., driving \pr{}s towards a terminal state which is completion or abandonment.  We received positive feedback (comments from developers) and observed that intended actions are taking place on the \pr{}s. To provide a glimpse, we list some of the quotes that we received from the developers that are appropriate to discuss in the context of this paper. On one of the \pr{}s, a developer said 

\begin{quote}
    \textit{``I agree. Making a few more changes and pushing this \pr{} through! Thanks for the notification!''}
\end{quote}
We then saw this developer acting on this \pr{} by pinging the reviewers and driving this \pr{} towards the completion within eight minutes. 

In another \pr{}, the developer first replied to the Nudge notification saying
\begin{quote}
\textit{``The pipeline is failing and blocking this check-in. Followed up with an ICM incident and completed the \pr{}!''} 
\end{quote}
Then, within a day, the \pr{} was abandoned. 
Thus, Nudge is not just about merging approved pull requests quicker,
but also about pushing \pr{}s to a terminal state, including abandonment,
and in this way maintaining repository hygiene.

For Nudge-LT, we also received feedback that says the notification is not useful because it is blocked by a reviewer. 
For example:
\begin{quote}
    \textit{``The comment does not add any value to me personally because I already know that the \pr{} I've authored has been open for a long time. It is not me who is blocking this but the reviewer''}
\end{quote}

Similarly, For Nudge-LT we see comments about why notification is considered not useful in cases where the  author interacted with the \pr{} recently by resolving a comment or pushing a new commit,
which we nevertheless ended up nudging because the lifetime of the \pr{} was long.
One such comment comes from a developer who she says 
\begin{quote}
    \textit{``I just resolved the comments on this \pr{} yesterday. I know about this one being pending for a while. This is not helpful!''}
\end{quote}

Both cases were in fact addressed by the actor and activity detection mechanisms of Nudge-FULL.


\subsubsection{Nudge-FULL user feedback}
Consistent with the many positive notification resolutions (Section~\ref{sec:resolution}),
many users were positive about the actor identification and activity detection enhancements. 
While there were some differences on how long the service should hold itself back before sending a notification when an activity is seen (24 hours vs 48 hours), we generally received agreement about the usefulness of these features. When asked about determining change blockers and ``@ mentioning'' them in the notification thus eliminating an extra hop, users stated: 
\begin{quote}
\textit{``Yes it'll be nice for the tool to ping the reviewers instead of having the person do it.''}
\end{quote}
\begin{quote}
\textit{``Yes I think that's handy to notify specific people. I often see someone "waiting" on a PR for changes, but then forget to revisit and follow up after changes have been pushed.''}
\end{quote}

Another user indicated that the algorithm was very accurate in determining the change blocker for a \pr{} that he was working on:
\begin{quote}
\textit{``Change blocker was perfectly identified and notified for \pr{} 731796. You did my job!''}
\end{quote}

While the deployment of the activity detection and actor identification modules reduced the negative feedback significantly, there remain cases where the developers expressed their dislike towards the Nudge notifications. For example:

\begin{quote}
\textit{``This \pr{} is awaiting on another \pr{} due to a module-level dependency. Thanks for the reminder though!''}
\end{quote}

\begin{quote}
\textit{``I know what I am doing. This is not helpful.''}
\end{quote}

\begin{quote}
\textit{``I went on a vacation. I would have liked it if you knew that and did not nudge me''}
\end{quote}

Suggestions on how to address this feedback are discussed in Section~\ref{discusion}.

\subsection{RQ4: Nudge at Scale}

To assess the impact of scaling up to thousands or repositories, we report Microsoft's experiences with deploying Nudge in production.
The initial deployment of the \PaperTitle\ service on \NumRepos\ source code repositories 
and the observed efficiency gains and positive user feedback 
convinced Microsoft management to deploy Nudge beyond the original repositories.
Thus, we trained and deployed the ``Nudge-FULL'' configuration for 8,000 repositories.
From January 2021 until December 2021, the
Nudge service sent notifications on 210,000 \pr{}s authored by 40,000 unique developers.
This deployment corresponded to an increase by a factor of 50 of in the number of repositories
compared to the initial experiment.
This increase was easily handled by Nudge, thanks to the fact that scalability was a design consideration right from the start.

We could not perform A/B testing as on the deployment on \NumRepos\ repositories of the \PaperTitle\ service due to administrative and logistical reasons. However, we were able to collect two important metrics from the large-scale deployment: (1) the positive resolution percentage, and (2) the distribution of \pr{} that are completed within a working day, three days, and a week.

We found that 71.5\% of the 210,000 \PaperTitle\ notifications were resolved positively. This is close to the 73\% positive resolution percentage from the \PaperTitle\ service deployment on \NumRepos\  repositories. 
Similar to the small scale deployment, 
16.35\% of the \pr{}s took more than a week to close (formerly 18.47\%), 
and 83.65\% of the \pr{}s were closed within a week (formerly 81.53\%).
These numbers indicate that the findings from RQ1--RQ3 continue to hold true when deployed at the scale of thousands or repositories.

\ \\
\section{Discussion} \label{discusion}
In this paper, we presented Nudge, a service for improving software development velocity by accelerating \pr{} completion. Nudge leverages machine learning-based effort estimation, activity detection, and actor identification to provide precise notifications for overdue \pr{}s.
Our experiments on \NumNudgedPrs{} \pr{}s in \NumRepos{} repositories over a span of 18 months demonstrate
a reduction in completion time by over 60\% (from 197 hours on average to 77 hours)
and 73\% of the developers reacted positively to being \emph{nudged} 
 --- numbers that continued to be valid when we scaled up Nudge to thousands of repositories.
In this section, we reflect on these contributions, assess their limitations, consider design alternatives, and explore future implications of our findings.

 
\subsection{Explicit Completion Times}
In our current implementation, \pr{} completion time is an attribute internal to Nudge,
that is not shared with the \pr{} authors.
An alternative design would be to let the author use the predictor to get an estimate of how long it would take to close this \pr{}, which they then can use to set a deadline for the \pr{} completion.
We did not pursue this route, because doing this might adversely impact the \pr{}s: The prediction
might become a self-fulfilling prophecy causing unnecessary delay \cite{article-confirmation, inproceedings-conf-software}.
Also, the \pr{} process will become unnecessarily complicated since the author and reviewers might engage in a back-and-forth discussion to decide the deadline.
 

\subsection{Interruptions} 
Nudge uses the existing functionality in Azure DevOps to remind the actors by adding comments to the \pr{}. These comments would result in email notifications which can be addressed asynchronously. This lightweight workflow is no different from other notifications which are sent when a reviewer is added to the \pr{} or they add a comment to the PR. Therefore, given the asynchronous nature and also based on the survey results, we do not believe that Nudge causes significant interruption for the reviewers. 
Also, recall that Nudge sends only one Nudge notification per \pr{} to minimize repeated interruptions.

Nudge does not reduce the total effort needed to complete a pull request.
Instead, it warns developers that others are waiting for them, 
suggesting them to prioritize the work on a given pull request.
The cost of this for the nudged developer is that some other work (ongoing coding activities, opening a new pull request, responding to another pull request) is delayed, while the nudged pull request is moved forward.
With Nudge, developers can take an informed decision whether to work on the \pr{} in question sooner rather than later.
In this way, they not just optimize their own queue of tasks locally, but can take a bigger picture into account, reducing the number of developers who are waiting for them to take action.


\subsection{Code review quality}
In our work on Nudge, we have focused on the calendar time duration of code reviews since it is deterministic and observable. 
Furthermore, in an industrial context, such speed of code reviews is important because of time-bound product release life cycles. In case of bugs and incidents, faster code reviews can help with faster resolution of bugs and quicker service restoration.

In this paper we have assumed that the total amount of effort in a \pr{} is not affected by Nudge:
tasks are moved earlier in time, but the nature of these tasks remains the same.
In line with that, we argue that the \emph{quality} of the reviews and code changes in nudged \pr{}s is not affected by Nudge.
Nevertheless, it could be the case that developers feel pressure based on nudges received,
and hence rush their work, and deliver lower quality.
On the other hand, it could also be that developers are able to deliver \emph{better} work,
since handling of the \pr{} takes place in a more confined time span, requiring fewer context switches, or context switches that are closer in time together.
We leave a rigorous investigation of the effect of nudging reviewers and developers on \pr{} quality as future work.

 
\subsection{Simplifying Lifetime Prediction}
An alternative to our learned lifetime prediction model is to work with a 
simple \emph{constant} model.
We explored this, as stated in Section \ref{RQ1}, by taking the mean of the \pr{} lifetime as the estimated lifetime for all \pr{}s in the population.
While simpler, such a constant approach suffers from the following problems:
\begin{enumerate}
    \item
    Nudge has been designed to be operationalized on tens of thousands of repositories,
    with different characteristics, processes/practices, and ever-changing dynamics.
    Thus, even a `constant' model is likely to require different settings across repositories
    and periodic re-calibration.
    \item
    The constant model will underestimate complex pull requests, yet overestimate simple ones.
    This may undermine the confidence in Nudge's notifications
    \item
    We conducted informal, small-scale user studies by showing the users the notifications and simulating the timing of the notifications of constant and actual Nudge-LT models. 
    Developers are inclined toward a model that adapts to changing workloads (dynamic), customized by user profiles or history, and that considers the size or complexity of the pull requests.
\end{enumerate}

\subsection{Addressing Nudge Limitations}

20\% of the Nudge notifications (882/4081) received an explicit \emph{won't fix} mark from the developers.
We recognize the following reasons, together with a potential way to address them.

First, a pull request may be blocked by the progress on another pull request.
Presently we do not take such inter-pull-request dependencies into account.
A possible next step is to scan pull requests for other pull requests mentioned in their discussions, 
and to consider such dependencies when nudging,
putting, e.g., more emphasis on blocking pull requests, and postponing nudging blocked pull requests until they are unblocked.

Second, while we have some level of detection to understand if a user is away, it is limited to detecting weekends and popular public holidays only, at this point. Future work includes incorporating an algorithm that looks at other data sources to detect and predict when a user will be away and account for that in the Nudge notifications.

Lastly, the Nudge system, at this point, does not ``learn'' based on user feedback. If a user passes negative feedback, Nudge does not use that information to pass that back to the model and adjust the parameters. Accounting for the user feedback, structuring it so that Nudge could leverage it, and determining the opportune moment to send the Nudge notifications are possible ways to further enrich Nudge.

\subsection{Threats to validity}
\subsubsection{Internal validity} Our qualitative analysis was conducted by reaching out to the developers via Microsoft Teams. 
None of the interviewers knew the people that were reached out or worked with them before. We purposefully avoided deploying Nudge on repositories that are under the same organization as any of the researchers involved in this work. As Microsoft is a large company and most of the users of the Nudge service are organizationally distant from the people involved in building Nudge, the risk of response bias is minimal. However, there remains a chance that respondents may be positive about the system because they want to make the developers of Nudge, who are from the same company happy. Lastly, for the error estimation of the machine learning models, we have used a single run of the ten-fold cross-validation. Using repeated cross-validation can result in a more accurate estimation of the performance of machine learning models.
 
\subsubsection{External validity} Depending on data availability and API usage policies, the Nudge model can be operationalized on other popular git-based source control systems like GitHub, GitLab, BitBucket, etc. However, the coefficients or the factors that impact the completion time of the \pr{}s, change blockers, etc may vary in those systems. Careful analysis of large samples of open-source data has to be performed before the Nudge model is deployed on systems like GitHub.
Some of the implementation details such as the heuristics used for identifying non-human actors will need to be adapted depending on the context.
Similarly, in the current implementation, we remove the 48 hours period corresponding to the weekend while computing \pr{} completion time, yet this may not be applicable to open-source projects.

The empirical analysis, design and deployment, evaluation, and feedback collection have been conducted specifically in the context of Microsoft. Given that Microsoft is one of the world’s largest concentration of developers and developers at Microsoft use a very diverse set of tools, frameworks, and programming languages, our research, and the Nudge system will have broader applicability. However, at this point, the results are not verified in the context of other organizations or the open-source community.

\section{Conclusion}
\prc{} is a key part of the collaborative software development process.
In this paper, we presented Nudge, a service for improving software development velocity by accelerating \pr{} completion. Nudge leverages machine learning-based effort estimation, activity detection, and actor identification to provide precise notifications for overdue \pr{}s. 
To make the notifications actionable, Nudge infers the actor, the \pr{} author or its reviewer(s), who is delaying the \pr{} completion.

We have conducted a large-scale deployment of Nudge at Microsoft where it has been used to \textit{nudge} over \NumNudgedPrs\ \pr{}s, over a span of 18 months, in \NumRepos\ repositories.
We have also conducted a qualitative and quantitative user study to assess the efficacy of the Nudge algorithm. 
Our findings include that 73\% of the notifications by Nudge have been positively acknowledged by the users. Further, we have observed a significant reduction in completion time, by over 60\% on average, for \pr{}s which were \textit{nudged}.

We further scaled out \PaperTitle\ to 8500 repositories at Microsoft and presented results from the large-scale deployment. We observe that \PaperTitle\ service was able to retain a good positive resolution percentage (71.5\%) similar to the deployment on \NumRepos\ repositories (73\%). We also observe that 83.65\% of the nudged \pr{}s were completed within a week similar to the deployment on \NumRepos\ repositories (81.53\%).

At the time of writing, the results reported in this paper have been the reason for Microsoft to explore adopting Nudge to a wider set of repositories.
Though culturally very different from Microsoft systems, we also believe Nudge-like functionality could be beneficial to repositories of many open source systems.
From a research perspective, we see future research in the areas of measuring the impact of shorter or longer reviewing cycles on reviewing quality, 
refining the pull request lifetime prediction models, 
taking inter-repository dependencies into account when nudging,
and estimating reviewer availability to make nudges as meaningful as possible.
\section*{Acknowledgements}
We would like to thank Rahul Kumar, Tom Zimmermann, B. Ashok, Suhas Shanbhogue, and Mei Nagappan for all their help with this work, and the anonymous reviewers for their valuable feedback.

\bibliographystyle{ACM-Reference-Format}
\bibliography{Paper}

\end{document}